\documentclass[graybox]{svmult}

\usepackage{mathptmx}       
\usepackage{helvet}         
\usepackage{courier}        
\usepackage{type1cm}        
\usepackage{makeidx}         
\usepackage{graphicx}        
\usepackage{multicol}        
\usepackage[bottom]{footmisc}

\usepackage{amsmath}
\usepackage{float}
\usepackage{subfigure}
\usepackage[numbers,compress]{natbib}

\newcommand{\indep}{\rotatebox[origin=c]{90}{$\models$}}

\begin{document}

\title*{Causal Queries from Observational Data in Biological Systems via Bayesian Networks: An Empirical Study in Small Networks}
\titlerunning{Causal Gene Network Inference with Bayesian Networks}
\author{Alex White and Matthieu Vignes}
\institute{A. White \& M. Vignes \at Institute of Fundamental Sciences; S. Marsland \at School of Engineering and Advanced Technology (SEAT); all \at Massey University, Private Bag 11 222, Palmerston North, 4442, New Zealand, \email{m.vignes@massey.ac.nz}}
%
%
\maketitle

\abstract*{Biological networks are a very convenient modelling and visualisation tool to discover knowledge from modern high-throughput genomics and post-genomics data sets. Indeed, biological entities are not isolated, but are components of complex multi-level systems. We go one step further and advocate for the consideration of causal representations of the interactions in living systems. We present the causal formalism and bring it out in the context of biological networks, when the data is observational. We also discuss its ability to decipher the causal information flow as observed in gene expression. We also illustrate our exploration by experiments on small simulated networks as well as on a real biological data set.}

\abstract{Biological networks are a very convenient modelling and visualisation tool to discover knowledge from modern high-throughput genomics and post-genomics data sets. Indeed, biological entities are not isolated, but are components of complex multi-level systems. We go one step further and advocate for the consideration of causal representations of the interactions in living systems. We present the causal formalism and bring it out in the context of biological networks, when the data is observational. We also discuss its ability to decipher the causal information flow as observed in gene expression. We also illustrate our exploration by experiments on small simulated networks as well as on a real biological data set.}


\section{Introduction}

Throughout their lifetime, organisms express their genetic program, i.e. the instruction manual for molecular actions in every cell. The products of the expression of this program are messenger RNA (mRNA); the blueprints to produce proteins, the cornerstones of the living world. The diversity of shapes and the fate of cells is a result of different readings of the genetic material, probably because of environmental factors, but also because of epigenetic organisational capacities. The genetic material appears regulated to produce what the organism needs in a specific situation. We now have access to rich genomics data sets. We see them as instantaneous images of cell activity from varied angles, through different filters. Patterns of action of living organisms can be deciphered using adequate methods from these data.
Probabilistic (in)dependence relationships can most certainly be extracted from observational data, but we want to go one step further. If the data is generated from a causal structure, what clues can be found in the data to help uncover this hidden structure without resorting to intervention? Intervention is not always practical, nor ethical. In our biological network study context, even though modern techniques allow scientists to intervene on the system, e.g. knocking out one gene or blocking a molecular pathway, they are still expensive experiments. We aim to investigate which kinds of causal links can (and which cannot) be discovered from observational data, defined by biologists as wild-type and steady-state.

In short, this chapter is reviewing causal reconstruction in gene network from observational data. More specifically, it blends a presentatioan application on a real network n of Systems Biology elements, methods for gene network reconstruction and concepts in statistical causality in Section~\ref{GRNcaus}. This section narrows down the type of system under scrutiny, namely gene regulatory networks, although our work could apply to more detailed representation of biological systems if the data were available. Section~\ref{sectionBN} introduces, in detail, the class of method we chose to focus on: Bayesian networks. Despite being designed to decipher probabilistic (in)dependencies in the data, we test its capacity to learn causality from observational data. This is the purpose of the preliminary experiments on small networks and an application on a real network in Section~\ref{expe}. All simulations were performed with the aid of the \texttt{R} package \texttt{bnlearn}~\cite{scutari2009, bnR}. We conclude with a short discussions in Section~\ref{conclu}.


\section{Biological networks and causality}
\label{GRNcaus}

Biological system inner organisation is at the origin, and is the essence of the diversity of life we observe around us, while allowing organisms to adapt to their environments. They have received growing attention for the last 20 years~\cite{edwards1999, kitano2002, noble2006}, and are complex both in terms of their many different kinds of interacting elements that they constitute; and in terms of the diversity of the responses that they drive. They can be understood at different levels: from the interaction of species within an ecosystem, to the finer molecular grain through interactions between cells or the communications between organs. We focus on the latter situation in the present work.

\paragraph{Networks for biological systems} 

Biological networks offer an integrated and interpretable tool for biological systems knowledge modelling and visualisation. Nodes in these networks stand for biological entities (e.g. genes, proteins), whilst edges encode interactions between these entities \cite{barabasi2004}. 
The definition of the biological entities is far from trivial - for example the definition of a gene is not unique~\cite{gericke2007, pennisi2007}. In our ``small world'' modelling representation of the reality~\cite{mcelreath2015}, we use simplified definitions of these biological elements, and coin them variables in our system. Some nodes can represent a phenotype of interest: e.g. the yield or level of resistance to a disease or to an adverse environmental factor ~\cite{scutari2014}. 
An edge, on the other hand, has a more diffuse meaning: ranging from a vague association of the measurements linked to the nodes, to a functional dependency between them driven by an explicit mechanism. 
We will be primarily interested in directed edges; we aim to go beyond co-expression or association networks~\cite{zhang2005, tenenhaus2010, rau2015}. 
In the real biological world, mechanisms exist which trigger pathways within the system of a living organism~\cite{monod1960} and sometimes between living organisms~\cite{hentschel2000, dupont2015}. On the analytical side of this, we aim to providing a meaningful representation of these interactions, not to say regulations between the components in the system under study. We will use \emph{directed acyclic graphs}~\cite{causality}, although it is known that gene networks do indeed contain feedback cycles~\cite{mangan2003, zabet2011, shojaie2014}. Solutions exists to consider feedback cycles in the Bayesian network framework. For example, time can be used to unfold the successive regulations taking place between the elements of a cycle~\cite{ghahramani1998, friedman1998, husmeier2003}. When a time-series is not available, some work detail the existence of feedback cycles as a super graph associated to a family of distributions instead of a unique factorised distribution~\cite{tulupyev2005}. This is essentially inspired by an inversion of the technique used to condense (aka contract) all strongly connected components (which contain cycles) into a single composite node~\cite{harary1965}. Notice that each distribution would then correspond to different causal assumptions which would then need be tested using additional data. In \cite{lacerda2008}, the authors give consistency results for an algorithm to recover the true distribution of non-Gaussian observational data generated from a cyclic structure.

If one is interested in representing a fine grain of biological details, interconnected networks depicting different types of biological entities can be used. A small example can be found in Figure 1 of Chapter 1 of the present book. Some apparent relationships between genes and phenotypes are often the result of direct regulatory links between different sorts of biological entities. We focus on this gene regulatory network (GRN) with the assumption that it can be deciphered from transcript levels.

\paragraph{Data and reconstruction methods} 
Modern high-throughput technologies such as RNA sequencing allow the researcher to have access to the simultaneous levels of all transcripts in a biological sample. They provide a partial snapshot of the state of the cell. A first hindrance in analysing such data set is related to their inherent noise and high-dimensionality~\cite{quackenbush2007}; much more variables are observed and the low number of samples makes the reconstruction task very challenging. A vast number of methods were developed to circumvent this difficulty with difference performance guarantees~\cite{buhlmann2011, verzelen2012, giraud2014}. A second obstacle is concerned with the non-linearity of the relationships~\cite{oates2014}, sometimes immersed in intricate temporal responses~\cite{shojaie2010, rau2010, marchand2014}. Another feature of modern biological data sets is that of missing observations, either due to technical fault or because all relevant variables cannot be monitored~\cite{chandrasekaran2012}; adequate techniques need be implemented to deal with this~\cite{blanchet2009, colombo2012, fusi2012, sadeh2013}. We will explicitly note the difficulty of modelling feedback cycles with traditional (yet advanced) analysis methods~\cite{mooij2011}.

Many review papers can now be found in bioinformatics literature on gene network reconstruction~\cite{dejong2004, markowetz2007, lee2009, emmert-streib2012}, and also in the statistical literature~\cite{maathuis2009, colombo2012, oates2012, fu2013}. 
Other papers generally compare their relative merits and pitfalls~\cite{werhli2006, altay2010, emmert-streib2010, marbach2010}. In this vein, the Dialogue for Reverse Engineering Assessment and Methods (DREAM) project, over the last decade, has also ran several challenges related to network reconstruction~\cite{marbach2012, meyer2014, hill2016}. Lessons were learned about the power of combining complementary reconstruction methods~\cite{marbach2012, allouche2013}, and such methods showed great performance in reconstructing real networks from genuine data sets related to cancer~\cite{bontempi2011, engelmann2015}, protein signaling~\cite{sachs2005, ness2016}, or to  fitness-related data~\cite{gagneur2013}. 
When only observational data is available, bounds on causal effects can be retrieved (see the discussion on EoC or CoE below as well) and verified by means of intervention (knock-out) experiments~\cite{maathuis2010, taruttis2015}. 
 In no instance are we pretending to be exhaustive, since this research area is vast and constantly expanding. Our focus is to test the ability of methods to decipher causal links from steady-state observational data. For example, we ignore single-cell data \cite{sachs2005} and gene expression time-series modelling \cite{oates2014, michailidis2013}. Neither do we consider the addition/integration of biological knowledge or of complimentary data sets~\cite{werhli2007} or a supervised framework~ \cite{mordelet2008}. Lastly, we do not cover purely interventional designs \cite{eberhardt2005, hauser2014, meinshausen2016}.
In some sense, our work is related to the work of~\cite{mooij2016} or that of~\cite{athey2016}, but we explore beyond the cause-effect pair of variables or the treatment effect. Borrowing ideas developed in any of these paradigms would certainly be a winning strategy, if they revealed useful information in our transcript level steady-state context.

\begin{figure}
    \centering
    \includegraphics[width = 0.8\textwidth]{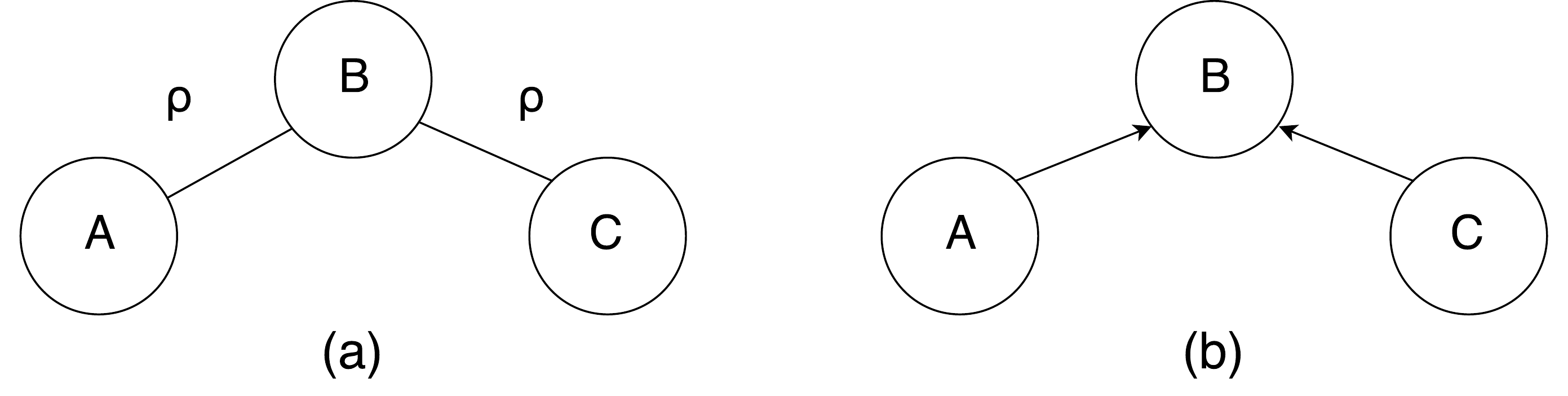}
    \caption{(a) Simple three-node network, and (b) reconstructed network from ideal data using the method in~\cite{opgen-rhein2007}. \label{figGGMpb}}
\end{figure}

The reader interested in a classified list of gene network reconstruction is encouraged to read through Sections 3 and 4 of Chapter 1 of the present book. We note here that  correlation-based methods do not lead to causal links between variables. An asymmetry (and hence directions on edges) can only be inferred by considering transcription factor genes as likely sources and other genes as likely targets (see also~\cite{chen2008} as a generalisation of~\cite{agrawal2002} in that it identified transcription factors). In a Gaussian graphical model setting, an attempt to infer directions was made in~\cite{opgen-rhein2007} using comparisons between (standardised) partial variances and the notions of relative endogeneous/exogeneous nodes.
The rationale of this distinction is that nodes with a higher level of remaining variance are more likely to be source of a directed edge, while those nodes with lower partial variance are more likely to be targets. The simple counterexample in the Box below and in Figure~\ref{figGGMpb} shows how directions can be created artificially. En route to causality, structural equation modelling (SEM, \cite{xiong2004}) assumes that the expression of each gene can be `programmed' as a linear combination of; every other gene expression (for non-zero path coefficients), all other factors, and of a noise term. The last method we mention here is a class of probabilistic graphical model, namely \emph{Bayesian Networks}. We postpone their detailed presentation until Section~\ref{sectionBN}. To the best of our knowledge, Bayesian Networks were first used for GRN reconstruction by~\cite{friedman2000}, and their ability to orient some edges and not others is well established~\cite{causality, spirtes2005, pearl2009}.

\noindent \fbox{
\begin{minipage}{0.95\textwidth}
We consider here the graph in Figure~\ref{figGGMpb} (a), which corresponds to a covariance matrix defined as
$$ M = \begin{pmatrix}
1 & \rho & \rho^2 \\
\rho & 1 & \rho \\
\rho^2 & \rho & 1
\end{pmatrix} $$
Now, assuming enough data is available, the maximum likelihood estimate of the concentration matrix, the inverse of the covariance matrix, which specifies direct relationship in a Gaussian graphical setting, is attained:
$$ K = \frac{1}{1-\rho^2} \begin{pmatrix}
1 & -\rho & 0 \\
-\rho & 1+\rho^2 & -\rho \\
0 & -\rho & 1
\end{pmatrix} $$
If we follow the direction rule of the approximate causal algorithm of~\cite{opgen-rhein2007}, we obtain the DAG in Figure~\ref{figGGMpb} (b). In fact, the partial variances of variables $ A $ and $ C $ are then equal to $ 1 - \rho^2 $ and hence larger than that of node $ B $, equal to $ \frac{1 - \rho^2}{1+\rho^2} $. Notice that this is conditional on having the edges and the directions declared significant, so it depends on the value of the correlation $ \rho $ and on the sample size.
\end{minipage}
}

\paragraph{Causality}

Causality is the intrinsic connection through which one event (the cause) triggers another one (the effect) under certain given conditions. Issues about causality and inductive learning are often ascribed to have originated with Hume's work on deciphering human cognitive reasoning as a scientific empirical study~\cite{hume1738}: that causation exists between our own experiments rather than between material facts; something can be learned from data. Major statisticians introduced such concepts in statistical data analysis at the beginning of the 20$ ^{\text{th}} $ century~\cite{wright1921, neyman1923, fisher1925}. The modern era of causality started with considerations of randomized vs. non randomized experiments in \cite{rubin1974} working with observational data, a review paper linking counterfactual and philosophical considerations~\cite{holland1986}, and \cite{pearl2009} aimed at unifying the different mainstream causal approaches. Science dived into seeking the Effects of Causes (EoC): what is the difference in outcome between the result of applying a treatment, and what would have happened if the treatment had not been applied. The latter of the two questions is referred to as a counterfactual statement.
In this framework, the cause is (partly) responsible for the effect, whilst the effect is (partly) dependent on the cause. 
Note the addition of the word partly here to clearly stress the likely, but not absolutely certain, aspect of the causation mechanism: statistics are concerned with stochastic events, not deterministic laws. Moreover, one cause is not exclusive of other causes for one consequent effect - several causal factors can be the cause of the effect, and can have additive effects or differing interactions with each other.
A causal sequence is generated and determined by natural mechanisms, generally several of them. These can be read from a temporal relationship. However, when nature does not explicitly reveal its mechanisms, learning about causality from data is extremely challenging. \cite{wainer2014} took a simple example so as to disentangle the research question; whether pupils achieving good results at school are happy because of their performance, or it is their intrinsic happiness which makes them good at school. In such cases, we need some sort of intervention achieved via randomisation of subjects to test hypotheses.

Historically, the language of counterfactual reasoning~\cite{bottou2013} has been used to define causal relationships from conditional statements ("what if something was different in the past"): if event $ A $ had have occurred, then $ B $ would have also occurred. Although the use of the subjunctive mood suggests the antecedent is false, the theory encompasses true or false antecedents. Though not universal, e.g. in a decision theoretic framework~\cite{dawid2000}, the counterfactual inference mechanism is natural to compute the EoC. \cite{causality} essentially formalised the use of counterfactuals for statistical causation inference. The "$ B $ would have been $ b $, had $ A $ been $ a $" statement means that the $ A \to B $ relationship can be read as "if we replace the equation determining $ A $ by a constant value $ A = a $, then the solution of the equations for variable $ B $ will be $ B = b $". (formally $  A = a \rightarrow B = b $). This allows the author to define the \emph{do} operator and state the property of the modification of the joint distribution under intervention (see Section~\ref{intervention}).
Among the tools used in causal inference, we mention Instrumental Variables~\cite{tan2006}, Causal Trees~\cite{athey2016}, Path Analysis~\cite{wright1921}, Structural Equation Modeling~\cite{bollen1989, tarka2017}, G-computation~\cite{robins1987}, Bayesian Decision-Making inspired approaches~\cite{dawid2000}, more recent Kernel Methods~\cite{lopez-paz2015}, and Probabilistic Causality~\cite{suppes1970, eells1991} (Pearl argued that a confusion was first made with the statement that cause increases probability \cite{causality}, whereas some assumptions need be checked to link the two notions).
Applications of statistical causality range from Psychology~\cite{buchsbaum2012}, Epidemiology~\cite{greenland1999}, Social Research~\cite{verkuyten2002}, Economics~\cite{spirtes2005}, Project Management~\cite{cardenas2017}, Marketing~\cite{gupta2008}, Human~\cite{kleinberg2011} and Veterinary~\cite{martin2014} Medicine. 

We now integrate the concepts of causality into systems biology and gene networks.  We will not address the relationships from marker to phenotype, although the kind of relationship can be similar, the causality necessarily originates from the genome  \cite{wu2010}. \cite{frommlet2016} introduces high-dimensional techniques for Quantitative Trait Locus (QTL) detection in recent data sets in which gene-gene interactions are accounted for. The perspective is more a pragmatical one for breeding research to improve traits of interest, rather than to dissect the molecular regulations. eQTL causal networks could be considered if we had genomic data in addition to transcriptomics~\cite{rakitsch2016}.

\paragraph{Causality in networks}

The concept of causality in gene network reconstruction is about distinguishing ``direct'' regulatory relationships~\cite{brazhnik2002} between biological components of the network from ``association''. 
Causal inference methods for gene network reconstruction are at the cornerstone of biological knowledge discovery, and may indeed be useful in solving biomedical problems. Randomised trials~\cite{hu2014} in this context are highly impractical, very costly, and even unethical in most instances~\cite{guyon2010}.
Gene co-expression is arguably not enough~\cite{djordjevic2014}, and most of the time methods rely on time-series \cite{anjum2009, rau2010, shojaie2010, deng2012, krouk2013, dondelinger2013}, perturbation \cite{cai2013, krouk2013} (sometimes combined with observational data \cite{rau2013, monneret2017}), or genetical genomics \cite{liu2008, allouche2013, tasaki2015} experiments. We will instead focus purely on observational (aka steady-state wild-type) data. 
As an example, an application of predicting the phenotype accounting for gene interaction and obtaining ranges of causal effects from a theoretical point of view can be found in~\cite{maathuis2010} with the R \texttt{pcalg} companion package in~\cite{kalisch2012}. 

Our choice is to test Bayesian Networks (BNs) for their inherent ease of representation of a complex system, with an associated probabilistic distribution under some mild assumptions. 
\cite{gupta2008} proposed that Bayesian networks have a limited applicability in the reconstruction of causal relationships. Methods for BN inference are reviewed in~\cite{koski2012} and in~\cite{bnR}. Arrows in these graphical networks are more like lines on the sketch of a construction plan - they encode conditional independence relationships, which by definition are symmetrical. Bayesian network arrows need not represent causal relationships, in fact, not even 'influencial' relationships. It is for this reason that it is perfectly valid to use prognostic or diagnostic models; the former estimates the chances of developing an outcome, and the latter estimates the chances of having this outcome~ \cite{hendriksen2013}. This distinction is also that between Effect of Cause (EoC, e.g. '$ Y $ is in state $ y $. Will a change in $ X $ have an effect on $ Y $?')  or Cause of Effect (CoE, e.g. '$ Y $ has changed its value. Is it because $ X $ was altered?')\cite{dawid2016}. With the exception of courthouse business, most causal questions - at least in science - are concerned with EoC via adequate designs of experiments~\cite{sebastiani2011}. CoE questions are hindered by situations where appropriate confounding effects are difficult to disentangle, perhaps due to missing information (for example missing variables~\cite{sadeh2013}). In this case, typically, one would need strong prior knowledge, or to make strong assumptions to allow the access of bounds on ``Probability of Cause'' to be estimated. To return to similar problems in gene network reconstruction, the reader is directed to~\cite{buhlmann2014}.

\paragraph{Causal interpretation of Bayesian Networks via node intervention}

Directed Acyclic Graphs (DAGs) like the one in Figure~\ref{fig:DAGexample} encode conditional independence relationships between the variables at the vertices. For instance, in this graph, $ S $ and $ W $ are independent, conditional on $ G $ and $ R $. $ G $ and $ R $ however, are not independent nor conditionally independent on any other subset of nodes.
In a pure graph theoretic approach, directed edges (parent/child relations), should not go through any interpretation beyond these independence relationships. Score-based or independence learning algorithms seek graph structures and conditional probabilities which optimally fit the data. On the frequentist side, there is no reason so as to prefer a member of the equivalence class of a learned graph (see Section~\ref{sectionBN}). On the Bayesian side, prior beliefs can steer our confidence towards one structure or another. 
Regardless, one is often tempted to say that a directed edge $ X \to Y $ stands for some kind of causal dependence of $ X $ on $ Y $. \cite{PC} asserted that one can retrieve causal relationships from observational data by resorting to considering directed edges of the essential graph. On the other hand, \cite{koski2012} claimed that taking the reasoning from immoralities to a causal interpretation is a fallacy. 
Unless there is clear knowledge in the system, causal statements are often  acceptably left vague, and they can only be clarified with the vocabulary of interventions in DAGs~\cite{pearl2009}. This extends the meaning of the independencies encoded in classical graphical models to an enriched set of distributions under interventions on nodes. An \emph{intervention} is defined as an operator setting a variable to a specified distribution. The most classical intervention one can think of is to impose a Dirac distribution to one variable - thereby setting it to some fixed value. Other distributions can be 'set' to any node in a DAG.

From these considerations, we aim to verify whether learning algorithms could infer simulated causal relationships, and in which settings. In other words: what can and cannot be learned from observational data in terms of causality in a 'typical' biological system?


\section{Bayesian Networks as a framework for DAG inference}
\label{sectionBN}
    
Bayesian Networks are a now widely used and powerful tool for probabilistic modelling. In this section we introduce this kind of graphical models and discuss their important features.

\subsection{Graphs and DAGs}
    	A \textit{graph}, $G$, is a mathematical construct consisting of a pair, $\mathrm{<}V, E\mathrm{>}$, of a set of \textit{nodes} (also known as \textit{vertices}), $V$, and a set of \textit{edges}, $E$. Each edge itself consists of a pair of nodes representing the endpoints of that particular edge. For example, an edge from node $a$  to node $b$ would be denoted $(a, b)$.\\
    	A \textit{DAG}, or \textit{Directed Acyclic Graph}, is a graph whose edges are directed; i.e. an edge $(a, b)$ goes \textit{from a to b}, not both ways; and which contains no cycles, i.e. for any node $a\in V$, there exists no non-trivial path which both begins and ends at $a$. \\
		If we consider each node as representing a particular event, and each edge as a conditional dependence, we begin to see how a DAG may be used as a graphical model. For example, consider a DAG with node set \texttt{\{(F = You run out of fuel), (L = You're late)\}}, and edge set \texttt{\{(F, L)\}}. We then have a two-node DAG with one edge going from \texttt{F} to \texttt{L}.

\subsection{d-Separation}
		The importance and use of d-separation will become obvious later, for now we simply provide the definition. Judea Pearl, considered by some to be the forefather of modern causal ideologies, defines d-separation as follows \cite[p.16-17]{causality}: 
		\begin{quotation}
		If $X$, $Y$, and $Z$ are disjoint subsets in a DAG $D$, then $Z$ is said to \textit{d-separate} $X$ from $Y$ ... if along every path between a node in $X$ and a node in $Y$ there is a node $w$ satisfying one of the following conditions;
		\begin{description}
		\item[1.]
 		$w$ has converging arrows, and none of $W$ or its descendants are in $Z$
		\item[2.]
 		$w$ does not have converging arrows and is in $Z$.
		\end{description}		
		\end{quotation}
		
\subsection{Probabilistic Independence}
		Two variables, $X$ and $Y$, are said to be \textit{probabilistically independent} given a third variable $Z$, denoted $(X \indep Y|Z)_P$, if 
		\begin{equation}
		    P(X,Y | Z) = P(X | Z) P(Y | Z)
		\end{equation} 
		Where 
		\begin{equation}
		    P(A|B) := \frac{P(A,B)}{P(B)}, P(B) \neq 0
	    \end{equation}
    	\subsection{D-maps, i-maps, and Perfect Maps}
    	A DAG, $D$, is an independency map, or \textit{i-map}, of some probability distribution $P$ if and only if d-separation of nodes in $D$ implies probabilistic independence of the corresponding variables in $P$, i.e. \begin{equation}(X \indep Y|Z)_D \rightarrow (X \indep Y| Z)_P\end{equation}
    	Conversely, a DAG, $D$ is a dependency map, or \textit{d-map}, if and only if \begin{equation}(X \indep Y|Z)_D \leftarrow (X \indep Y|Z)_P\end{equation}\\ A DAG, $D$, is a \textit{perfect map} of $P$ if and only if it is both a d-map and an i-map of $P$.
    	
\subsection{The Markov Blanket}
    	The Markov blanket of a node, $X$, in a DAG, $D$, denoted $MB(X)$, is the minimal set of nodes conditioned on which $X$ becomes independent of the rest of the nodes in $D$.
    	\begin{figure}[H]
            \sidecaption 
            \includegraphics[scale=0.45]{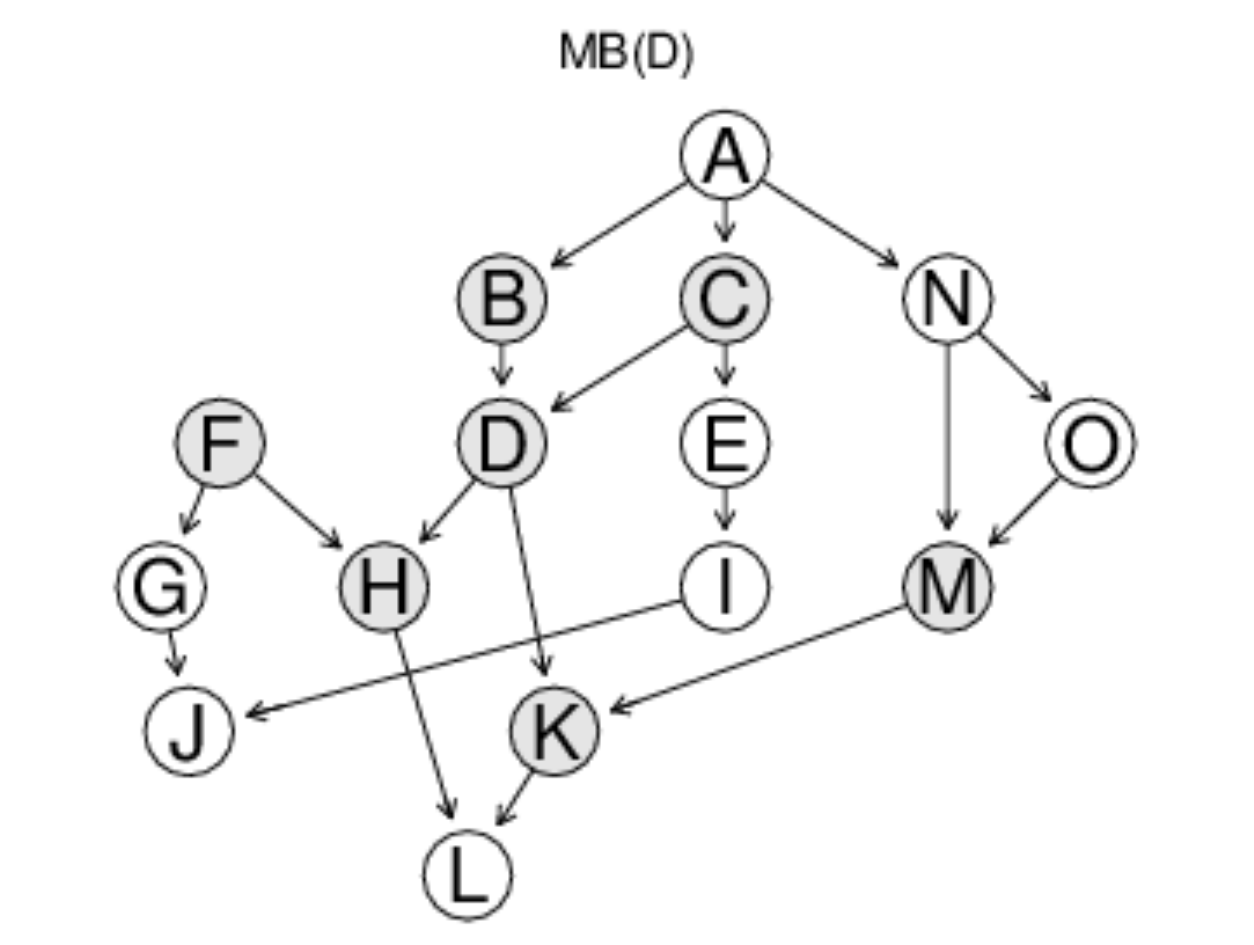}
            \caption{The Markov Blanket of Node D}
            \label{fig:MKD}
    	\end{figure}
        It can be shown \cite[p.120-121]{probReasoning} that if $D$ satisfies the Markov Condition -each node of the DAG is conditionally independent of its non-descendants given its parent variables-, $MB(X)$ consists of the parents, children, and children's parents of $X$. $MK(D)$ is illustrated in Figure \ref{fig:MKD}.
    	
\subsection{Bayesian Network definition}
    	A Bayesian Network (BN) is a minimal i-map of some probability distribution $P$; that is, a directed acyclic graph, $D$, whose node-wise graphical separations imply probabilistic independencies in $P$, and where any further edge removal from $D$ will result in loss of its i-mapness \cite[p.119]{probReasoning}.\\
	    A Bayesian Network, $B$, may be expressed as a pair $B = \langle D, \theta \rangle$ consisting of a DAG, $D$, and a set of parameters for the \textit{conditional} probabilities of the nodes in $D$, denoted $\theta$.
	    Many extensions to the Bayesian Network exist; for example, Object-oriented Bayesian networks \cite{koller1997} can be used to form compound models for more complex situations when some elementary blocks are repeated and hierarchically assembled.

    \paragraph{Why are they useful?}
    	
    	To begin, notice that in the definition for a BN, we only required $\theta$ to be the set of parameters for the \textit{conditional} probabilities for each node, not the entire joint distribution. This in itself simplifies expression of the full system. We also have that a BN satisfies the Markov Condition, and as such the maximum set of nodes that a node needs to be conditioned on is simply its Markov Blanket. This can drastically reduce the number of values required to fully specify the model. 

\subsection{Bayesian Network Learning: An Overview} \label{bnlearningsection}
    	
\begin{figure}[H]
    	    \sidecaption
            \includegraphics[scale=0.30]{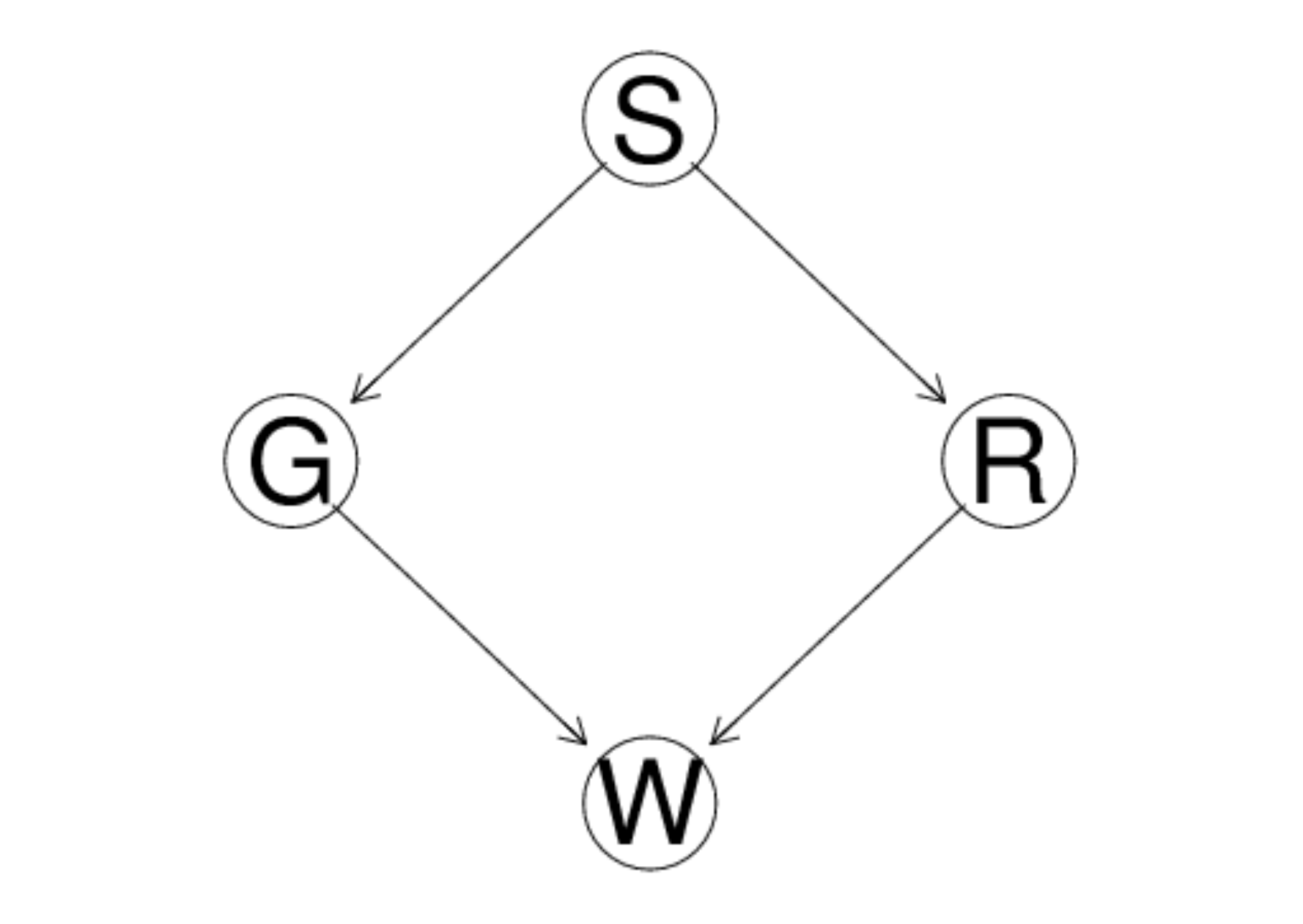}
  			\caption{An Example Bayesian Network}
           	\label{fig:DAGexample}
		\end{figure}

    	A sample from a Bayesian Network, $B$, consists of a realisation of each node in $B$. For example, consider the Bayesian Network in Figure~\ref{fig:DAGexample} as a classic example. 
		Let node $S$ denote the event that it is summer, node $R$ denote the event that rain is falling, node $G$ that your garden sprinkler is active, and $W$ that the grass is wet. A sample from this network could look something like: $$\{\texttt{summer, } \texttt{rain, } \neg\texttt{sprinkler, } \texttt{wet}\}$$
    	Many algorithms exist to reproduce either the structure or probability distribution associated with a BN from a set of such samples, and can be summarised as in the following sections.

\subsection{Structure Learning}
    	Learning the structure of a BN is a complex task. The number of possible acceptable edge configurations grows super-exponentially as the number of nodes increases, for two-node networks there are 3 possible valid DAG structures. For four-nodes there are $543$ structures, and for eight nodes there are $\approx 7.8 \times 10^{11} $ structures. For a reasonable twenty-node network there are $\approx2.345 \times 10^{72}$ possible valid edge configurations~\cite{Stephen}. This makes exploring the entire search space of plausible (and large enough to be useful) BNs an impractical exercise. Following are overviews, and examples, of the three main heuristic approaches intended to deal with this problem.\\

\paragraph{Constraint-Based Algorithms}
    	The first general approach is using \textit{constraint-based} algorithms, which are usually based on Pearl's \textit{IC} (Inductive Causation) algorithm \cite[p.50]{causality}. With these, we start with a fully saturated, undirected graph and use conditional independence tests such as the log-likelihood $G^2$ or a modified Pearson's $\chi^2$ test \cite[p.99]{bnR} to remove edges one-by-one. Another pass then adds direction to edges where adequate information is available (see Pearl's algorithm \cite[p.50]{causality}). Examples of constraint based algorithms include \textit{IC} \cite[p.50]{causality}, \textit{Incremental Association Markov blanket (IAMB)} \cite{IAMB}, and \textit{PC} \cite[p.84]{PC}.

\paragraph{Score-Based Algorithms}
    	The other common approach is one which considers and scores the network as a whole, rather than one edge at a time. Initialise any network of your choosing, then possible future networks are evaluated by providing them with a score; such as the \textit{Bayesian Dirichlet Equivalent uniform} (BDE), or the \textit{Bayesian Information Criterion} (BIC) score \cite[p.106]{bnR}. 

\paragraph{Hybrid Algorithms}
    	Perhaps most commonly used are hybrid algorithms, which, as the name would suggest, are a combination of the two previous approaches. For example, the Sparse Candidate algorithm \cite{SparseCandidate} uses tests for independence to restrict the search space for the subsequent score-based search. This general approach is known as the \textit{restrict-maximise} heuristic, and is implemented in \texttt{bnlearn} as \texttt{rsmax2}.

\subsection{Parameter Learning}
    	Once the structure of a network is known, the parameters to fit can be estimated relatively easily, using either; a frequentist maximum likelihood approach, whereby you view the global distribution as a function of the model parameters and do some optimisation over the data; or by Bayesian methods \cite[\S1.4]{bnR}, where you update a prior belief about the parameters in light of the data \cite[p.11]{bnR}. The first method, in the case of discrete Bayesian Networks, reduces to simply counting the occurrences of the data points relevant to each node and using the sample proportions to estimate the parameters. The Bayesian approach assigns a uniform prior over the entire conditional probability table, and then calculates the posterior distribution using the data present. Both methods are implemented in \texttt{bnlearn} in the \texttt{bn.fit} function.
        
\paragraph{Two Notes on Using Bayesian Methods}
Firstly, the use of Bayesian Methods may prove valuable when there is expert information available, as the uniform prior mentioned above does not necessarily need to be uniform - it may just as well represent the belief of an expert in the relevant field. 
To use the Bayesian approach in parameter learning, \texttt{method = 'bayes'} must be included in the call to \texttt{bn.fit}. This allows for specification of the prior distribution. The user has some control over the weighting of the prior distribution with respect to the data when calculating the posterior distribution by means of the \texttt{iss} (imaginary sample size) argument.

Secondly, a discussing remark must be made on the scalability of the Bayesian network approach. Continuous efforts are made to improve the network size which can be handled. Just over a decade ago, one could tackle a few hundred nodes in a sparse network~\cite{brown2004}. In the case of discrete data, the algorithm complexity is exponentially linked to the number of classes to represent factor variable levels. When the data is continuous, several choices are possible~cite{fu2005}, from prior discretisation to direct modelling, e.g. relying on kernels. In conjunction to discretisation considerations, the maximum number of parents allowed per node constrains the capacity of the algorithm to run on a given data set~\cite{vignes2011,qi2016}. Beyond computational smart decisions, theoretical studies characterise algorithm complexity~\cite{mengshoel2010, decampos2011} since~\cite{cooper1990}. This leads to the development of more efficient algorithms~\cite{handa2003, malone2011, adabor2015}. Sometimes, developments rely on parallel computation~\cite{nikolova2012, madsen2017}. Bayesian networks can be learnt with thousands of variables and few tens of thousands of edges using most current statistical packages. More specific applications can allow practitioners to perform computations with hundred of thousands of nodes, e.g. to perform variable selection in databases~\cite{thibault2009}, but this is not yet the case nor a need for biological networks, where a few tens of thousands of nodes is amply enough, and the limiting factor is then rather sample sizes.
    	
\subsection{Intervention} \label{intervention}

\begin{figure}[H]
     		    \sidecaption[t]
            	\includegraphics[scale=0.3]{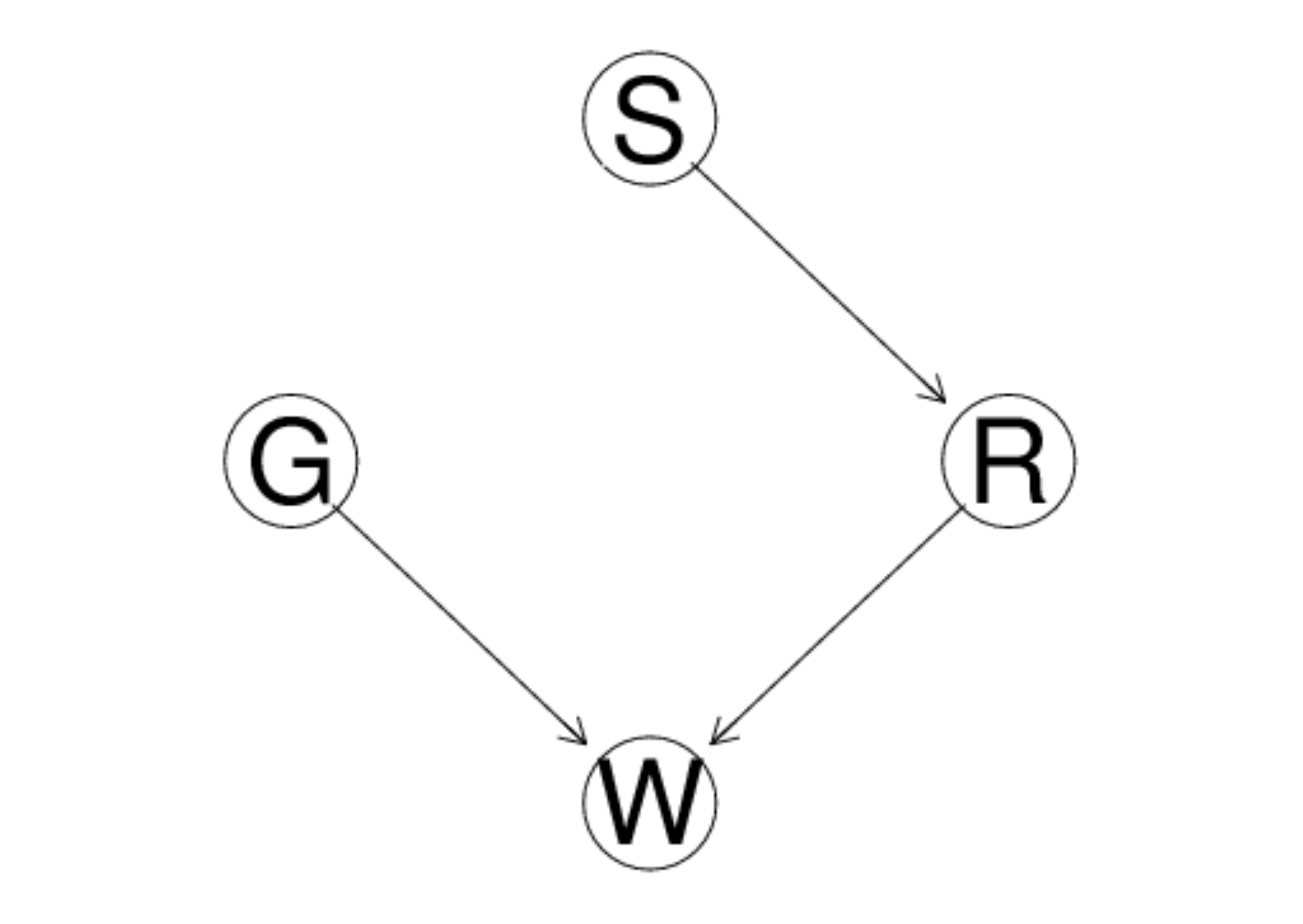}
  				\caption{The Bayesian Network from Figure \ref{fig:DAGexample} Under the Intervention $do(G=g)$}
    	       	\label{fig:InterventionalDAG}
			\end{figure}

    		Intervening on a network and setting the value of a node has a notably different effect than simply `observing' it. Consider again the DAG in Figure \ref{fig:DAGexample}, with semantics as defined in Section \ref{bnlearningsection}.\\
    		Were an external deity to intervene on the scenario and turn on the sprinkler, whether or not the sprinkler is on becomes independent of what season it is (as it has been forced on), and the DAG structure changes to that in Figure~\ref{fig:InterventionalDAG}.
			
			In this way, intervention renders a node independent of its parents, and the probability distribution underlying the DAG also gets altered. The modification (i) deletes the conditional probability of the nodes intervened on and (ii) sets their value when they are parents of other nodes in the relevant conditional probabilities. Notice that interventions do not necessarily fix the value of a node to a given value, but could impose a particular distribution.
			It should be fairly clear, then, that you cannot intervene on a set of samples, as they have already been drawn from a particular distribution which it is now too late to alter. 
            For example, if we have a set of samples from the DAG in Figure~ \ref{fig:DAGexample}, and we now want to know what would happen if we turned the sprinkler on (regardless of the season), we cannot do so without further experiment or more information about causality in the system.
            
Causal discovery in observational studies is mainly based on three axioms which bind the causality terminology to probabilistic distributions. The first one is the causal Markov condition. It states that once all direct causes of an event are known, the event is (conditionally) independent of its causal non-descendants. The second one is the faithfulness condition; it ensures that all dependencies observed in the data are structural, i.e. were generated by the structure of an underlying causal graph. These (in)dependencies are not the result of some cancelling combination of model parameters, an event which often has a zero probability measure with commonly used distributions~\cite{PC}. The last condition is termed causal sufficiency assumption; all the common causes of the measured variables are observed. When this is not the case, hidden variable methods can help~\cite{stegle2010}.

\subsection{Markov Equivalence}
			Consider a two-node network, $D$, with nodes $A$ and $B$, and one edge from $A$ to $B$, i.e. $A \longrightarrow B$. \, With a trivial application of Bayes' Rule, we observe:
			\begin{eqnarray} 
			P(A,B) &=  P(A)P(B|A) \label{AonB}\\ 
			&=  P(A)\frac{P(A|B)P(B)}{P(A)}\\
			&= P(B)P(A|B) \label{BonA}
			\end{eqnarray}
			By \ref{AonB} and \ref{BonA}, $P(A)P(B|A) = P(B)P(A|B)$ and the network is equivalent to $B \longrightarrow A$.

			Clearly, some visually non-identical networks can encode the same information and dependency structures, and samples from them will be indistinguishable. Such networks are known as \textit{Markov Equivalent} networks. For example \cite{he2013} used an MCMC computational approach to identify the equivalence class of a DAG. Calculating the values of $P(B)$ and $P(A|B)$ given $P(A)$ and $P(B|A)$ may be done as follows: \begin{equation}P(A|B)=\frac{P(B|A)P(A)}{P(B)}\end{equation}
			\begin{equation}P(B)=\sum_A P(B|A)P(A)\end{equation}


\section{Experimental setup: quality of a causal network reconstructed from observational data}
\label{expe}

    	The aim of our experiment is to investigate the similarity between the BN used to generate samples, and the BN learned from the data, as we varied the number of samples used in the process. This similarity was measured using the Structural Intervention Distance (SID)~\cite{SID}. We provide the code for our experiments at {\small \texttt{https://github.com/alexW335/BNCausalExperiments/blob/master/SID.R}}.
    	
\subsection{Method}

\begin{figure}[H]
    		    \sidecaption
                \includegraphics[scale=0.5]{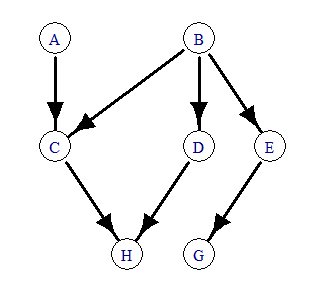}
  				\caption{Data Generating 7 node DAG Structure}
           		\label{fig:originalDAG}
			\end{figure}

    	To be able to more easily investigate the problem at hand, we decided to use a small, 7 node BN as shown in Figure \ref{fig:originalDAG}.
		
\begin{figure}
        \centering
            \subfigure[V-Structure]{%
            \label{fig:vstruct}%
            \includegraphics[height=0.75in]{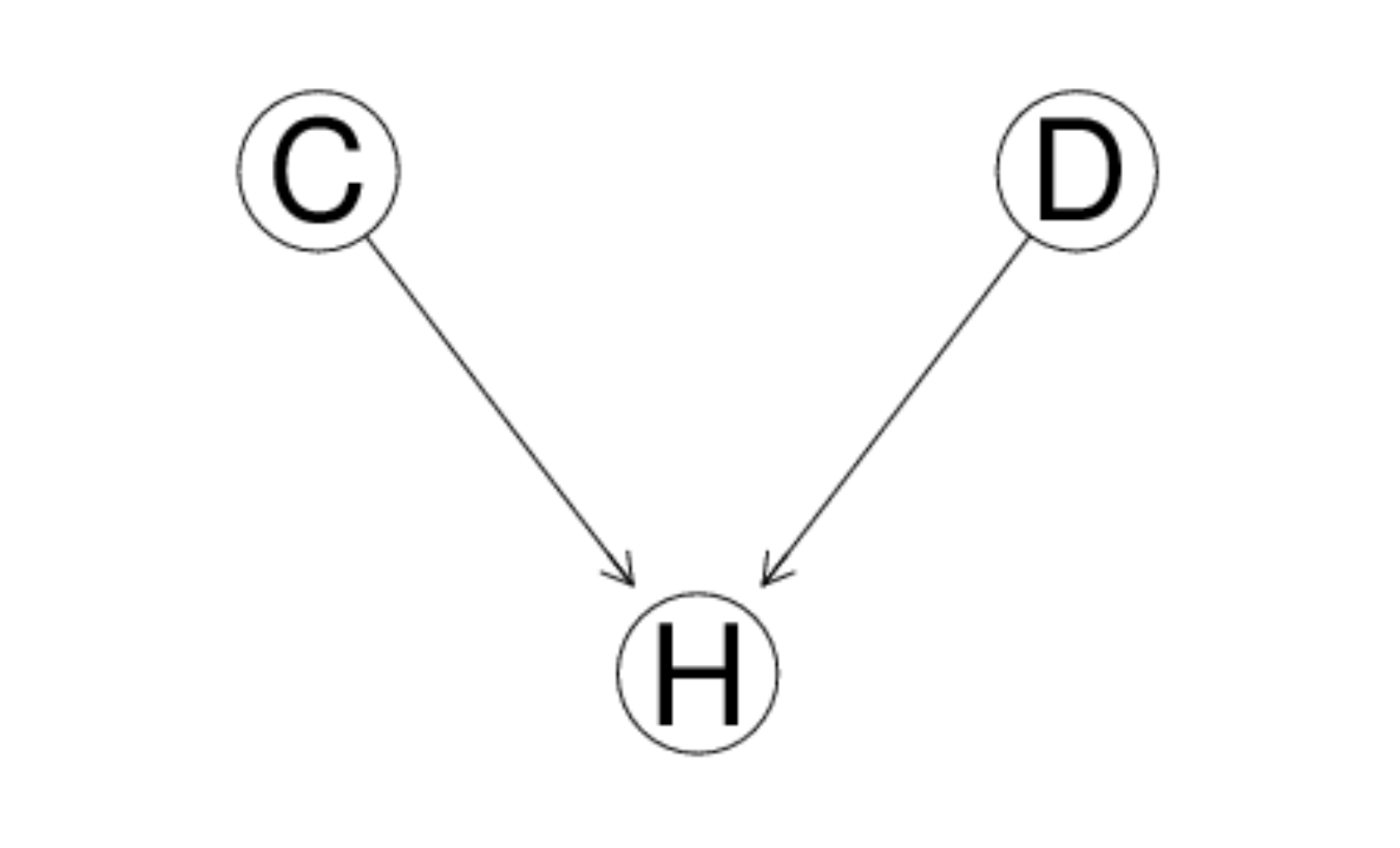}}%
            \qquad
            \subfigure[Chain]{%
            \label{fig:chain}%
            \includegraphics[height=0.75in]{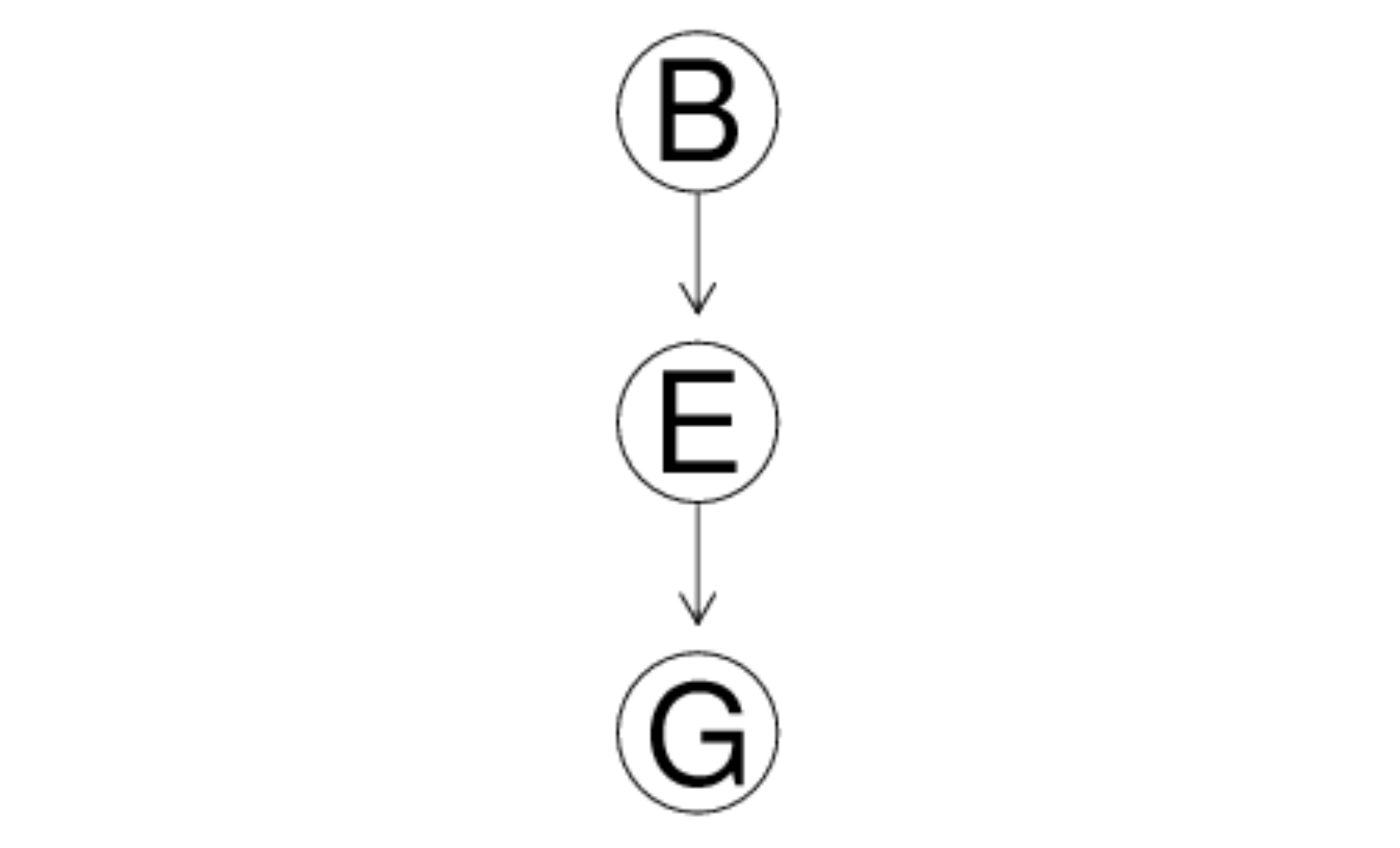}}%
            \qquad
            \subfigure[Fork]{%
            \label{fig:fork}%
            \includegraphics[height=0.75in]{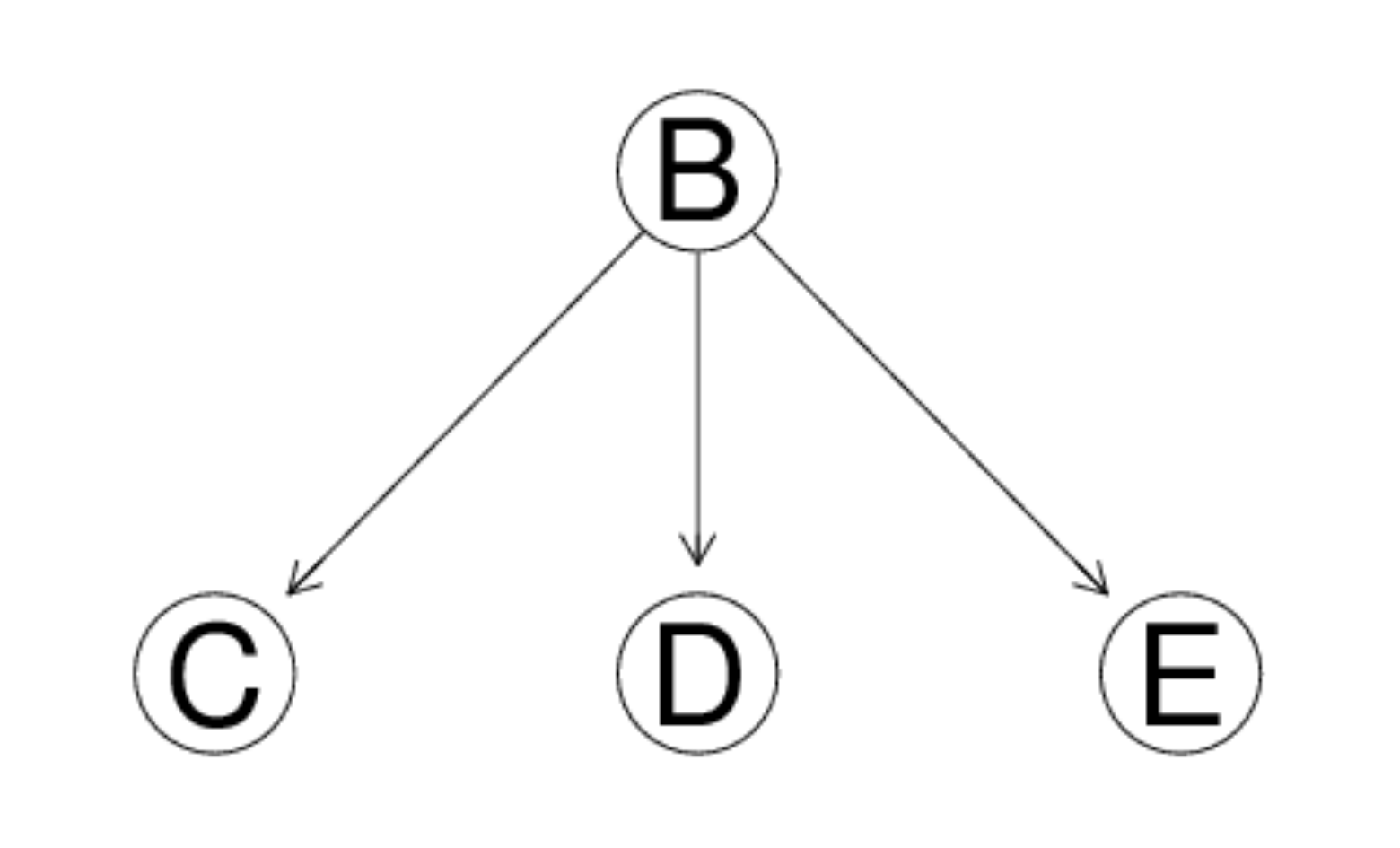}}%
        \caption{Common Structures of Interest \label{ExampleStructures}}
        \end{figure}

		This particular network was chosen as it possesses many substructures of particular interest when investigating causality, for example the v-structure (or collider) in Figure \ref{fig:vstruct}, the (causal) chain as shown in Figure \ref{fig:chain}, and the fork (or common cause) seen in Figure \ref{fig:fork}.
			
		The conditional probabilities associated with each node were chosen to show distinct separation in the data, e.g. $P(B=b|A=a)$ should be sufficiently different to $P(B=b|A=\neg a)$ for ease of DAG-edge recovery, i.e. we seek high model identifiability (related to faithfulness of the model). These conditional probabilities are available in the source code.
		
\paragraph{Data Generation}

    		Samples were generated on demand with the \texttt{R} function \texttt{rbn} \cite{bnR} from the package \texttt{bnlearn}.
    		
\paragraph{Network Reconstruction}

    		The two phase restrict-maximise heuristic was used to recover the DAG structure from the generated data. The semi-parametric $\chi^2$ test was used to check for conditional independencies, the Inter-Associative Markov Blanket (IAMB) algorithm~\cite{IAMB} was used in the restriction phase for locally minimising the search space, a straightforward hill-climbing algorithm was used for exploring the restricted space, and the Bayesian Information Criterion (BIC) score~\cite{BIC} was used as the score by which to compare the possible networks during the search.
    		
\paragraph{Initial Notes on BN Comparison}

    		Initially, two measures were considered for comparing the quality and similarity of the reconstructed Bayesian Network, $B^* = \langle D^*, \theta^* \rangle$, to the original Bayesian Network $B = \langle D, \theta \rangle$:
    		\begin{enumerate}
				\item \textbf{Graph Edit Distance}
 				The graph edit distance between $D$ and $D^*$, where one edit is either an edge \textit{addition}, \textit{deletion}, or \textit{reversal}. This compares the structural similarities of the DAGs, but fails to account for any differences in the underlying distribution.
 				
				\item \textbf{Kullback-Leibler (KL) Divergence} 
				KL Divergence from $B$ to $B^*$ may be used to compare the underlying probability distributions of the two BNs. As prior information about the structure of the original network is available in an experimental setting such as this, it is possible to condition the calculation of the KL Divergence on a particular node, say $D$; i.e. enumerate \begin{equation}\sum_{A}\sum_{B}\sum_{C}(P(D | A, B, C))\end{equation}  rather than \begin{equation}\sum_{A}\sum_{B}\sum_{C}\sum_{D}(P(A, B, C, D))\end{equation} Where $\sum_{X}\limits$ means sum over all possible states of $X$, e.g. \{$x$, $\neg x$\}.\\
 				This would give an idea of how accurately the probability distribution was being reproduced, and can be used to give some insight into the causal structure underlying the BN in question.
			\end{enumerate}
		    However, a few issues arise from using these measures. The KL Divergence implementation proves particularly difficult to generalise, and the conditioning on some node $D$ requires prior knowledge that the specific node selected would be the best node to look for down-flow effects. 

\paragraph{Bayesian Network Comparison with SID}
		    One metric was used to compare the two models - the Structural Intervention Distance \cite{SID}. This does not consider the probability distributions underlying the BNs in such a way that, for example, the Kullback-Leibler Divergence does; it rather contains information about the difference between both structures under the same intervention. This allows some information about the causal structure to be inferred, such as the location of v-structures and forks (See Figure \ref{ExampleStructures}). In a nutshell, SID measures the number of interventions which lead to different graphical structures between two DAGs. The smaller the SID, the closer the two DAG structures.

\subsection{Results}

\subsubsection{Seven-Node Network}

\begin{figure}[H]
           		\sidecaption
                 \includegraphics[scale=0.45]{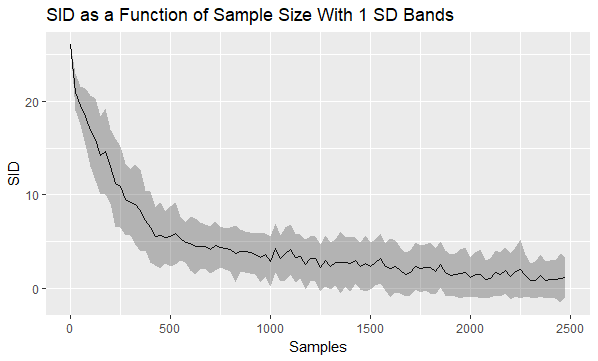}
  				\caption{Mean SID between the data generating Bayesian Network and a thousand BNs reconstructed from the samples, plotted as a function of the number of samples used to reconstruct the BN. A grey band of +-1 standard deviation for the mean SID is shown.}
           		\label{fig:SIDbySamples7}
			\end{figure}

		    As expected, the average SID appears to decrease as the number of samples used to generate the network increased, as per Figure \ref{fig:SIDbySamples7}. 
			
\begin{figure}[H]
           		\sidecaption
                 \includegraphics[scale=0.45]{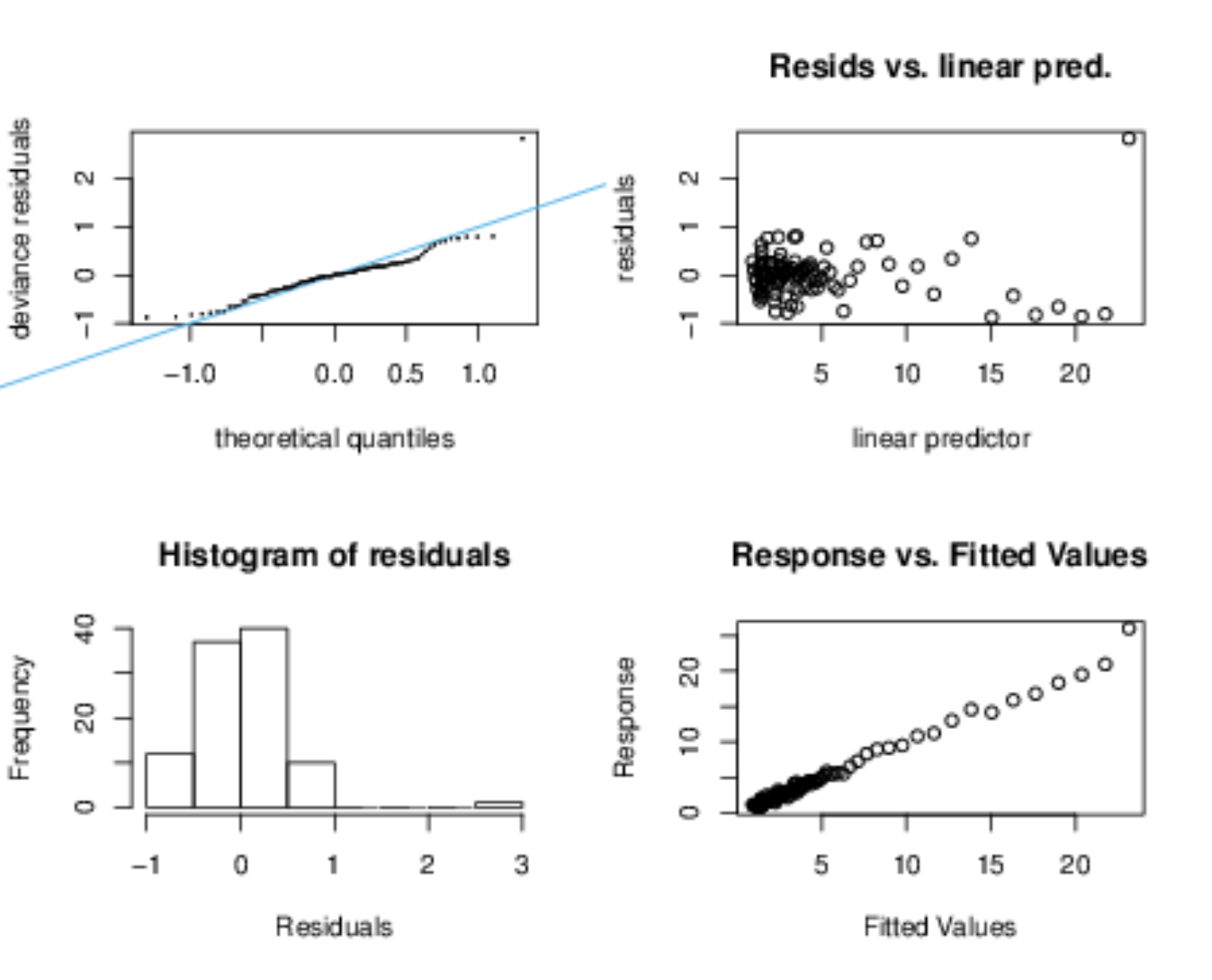}
  				\caption{Diagnostic plots for the GAM.}
           		\label{fig:GAMdiagPlot}
			\end{figure}

			The data shown in Figure \ref{fig:SIDbySamples7} can be well modelled using a Generalised Additive Model (GAM), though the usefulness of such a model is debatable. It is assumed that the SID should asymptotically approach zero (or is at least non-negative), so any model fitted should also exhibit this behaviour to prove itself a useful predictive tool. This aside, the diagnostic plots for the GAM may be found in Figure \ref{fig:GAMdiagPlot} to convince the reader that the model is at least naively fitting the observed data well.
			
\begin{figure}[H]
           		\sidecaption
                 \includegraphics[scale=0.45]{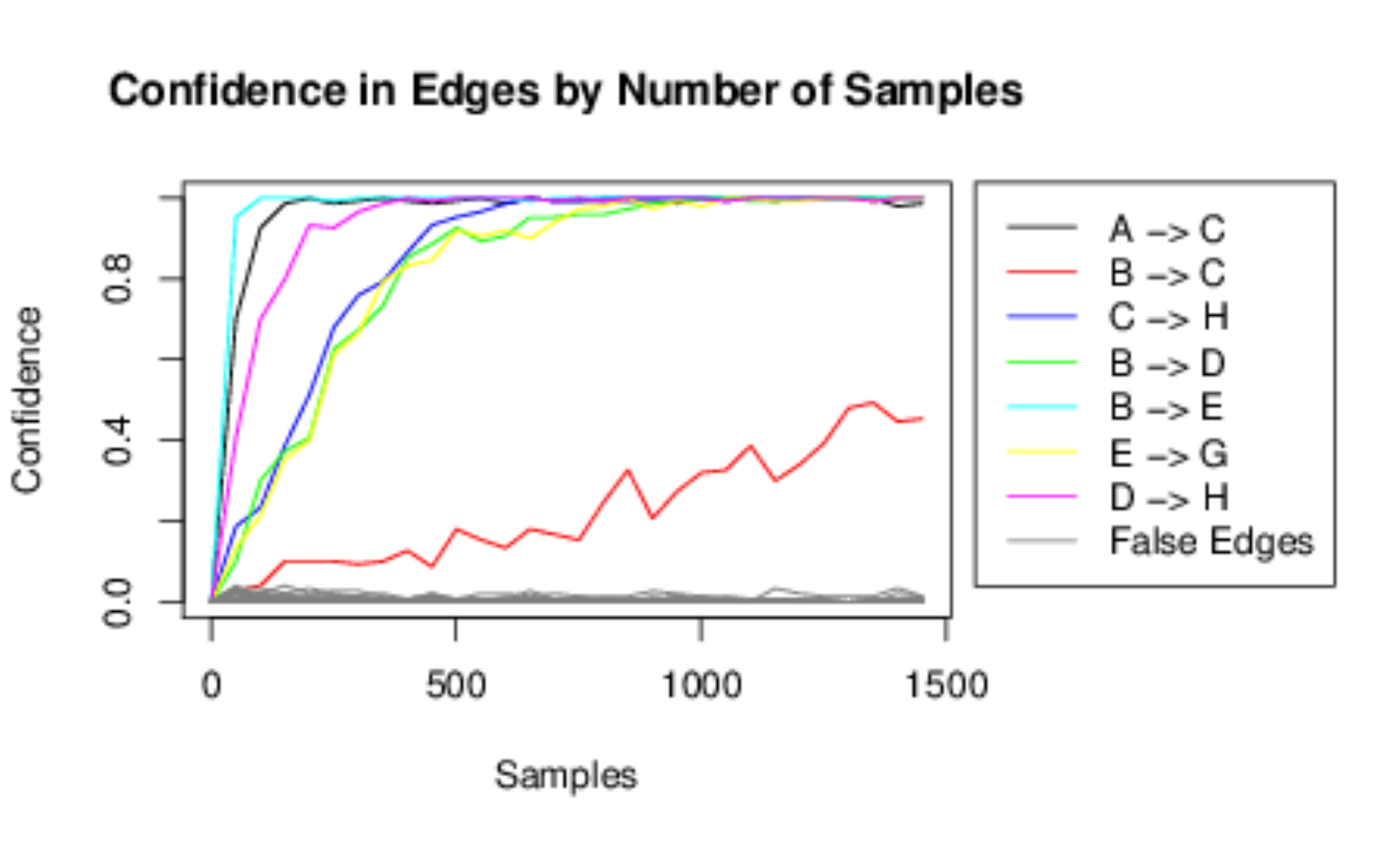}
  				\caption{Proportion of graphs generated from each number of samples which contain specific edges. The labelled edges are the ones which are known to be present in the actual BN.}
           		\label{fig:EdgeConfidence}
			\end{figure}

			Another factor of interest is how the confidence in the edges (both true edges, and those which should not appear) grows with the number of samples used to generate the network. This is plotted in Figure \ref{fig:EdgeConfidence} where the confidence at each level of sample count is expressed as the proportion of the thousand BNs generated in which each edge appears. As one would expect, the 7 true edges tend to increase in confidence as a function of the number of samples, and the remaining thirty five edges which are not present in the true BN appear to fluctuate randomly close to zero.
			
\begin{figure} \label{AverageGraphs}%
            \centering
            \subfigure[Average DAG Generated from 100 Samples]{%
            \label{fig:100sampleAV}%
            \includegraphics[height=1.5in]{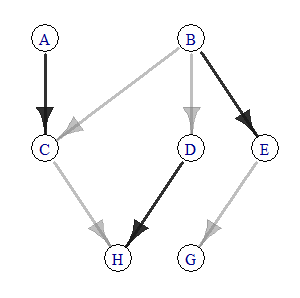}}%
            \qquad
            \subfigure[Average DAG Generated from 2500 Samples]{%
            \label{fig:2500sampleAV}%
            \includegraphics[height=1.5in]{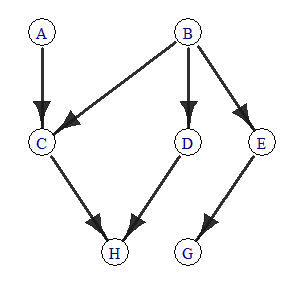}}%
            \caption{Average graphs generated from different numbers of samples. Note that the edge colouring has been passed through a three-stage threshold: edges appearing in less than 5\% of DAGs were not shown, those appearing in 5-50\% were shown in grey, and those in over 50\% were shown in black. }
            \end{figure}

			The same data used to produce Figure \ref{fig:EdgeConfidence} can be used to construct the average adjacency matrix, and thus the average DAG, produced at each sample count. The average DAG for 100 samples and 2500 samples are shown in Figures \ref{fig:100sampleAV} and \ref{fig:2500sampleAV} respectively.
			
\begin{figure}[H]
           		\sidecaption
                 \includegraphics[scale=0.45]{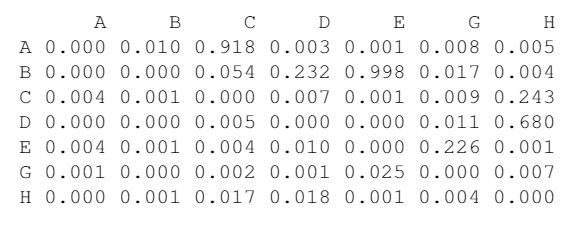}
  				\caption{The average adjacency matrix for DAGs generated from 100 samples. This corresponds to Figure \ref{fig:100sampleAV}.}
           		\label{fig:AdjMat}
			\end{figure}

            The adjacency matrices for the average DAGs provide more information than the simple graphical view, however they can be a little harder to decode as is evident in Figure \ref{fig:AdjMat}.
			
			\subsubsection{Four-Node Network}
\begin{figure}[H]
			\centering
           		\sidecaption
                 \includegraphics[scale=0.3]{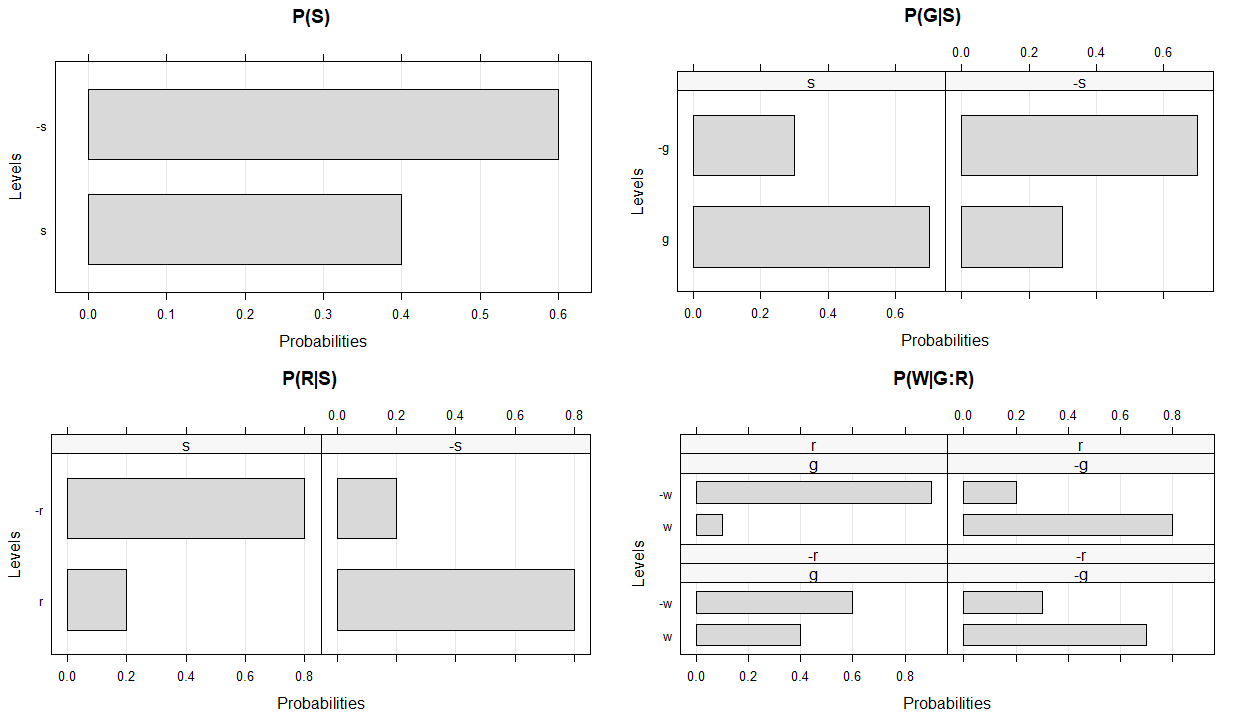}
  				\caption{The conditional probability tables for the nodes of the BN used in the second experiment.}
           		\label{fig:CondProb}
			\end{figure}

			As well as the illustrative seven-node network, the experiments were run on the Bayesian Network whose topology is shown in Figure~\ref{fig:DAGexample}, with conditional probabilities as represented in Figure~\ref{fig:CondProb}. Note that this network is unrelated to the similarly structured weather modelling Sectionnetwork of Section~\ref{bnlearningsection}.
			
\begin{figure}[H]
			\centering
           		\sidecaption
                 \includegraphics[scale=0.3]{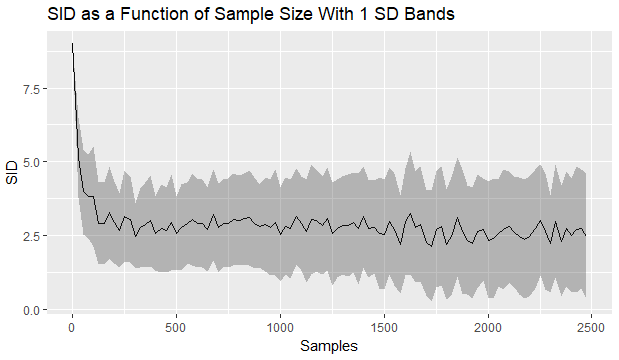}
  				\caption{Mean SID between the data generating Bayesian Network and a thousand BNs reconstructed from the samples, plotted as a function of the number of samples used to reconstruct the BN. A grey band of +-1 standard deviation for the mean SID is shown.}
           		\label{fig:SIDSamplesFourNode}
			\end{figure}

			This yielded a curve of SID by Samples as shown in Figure \ref{fig:SIDSamplesFourNode}, displaying markedly more variability than the seven-node case. 
			
\begin{figure}[H]
           		\sidecaption
                 \includegraphics[scale=0.45]{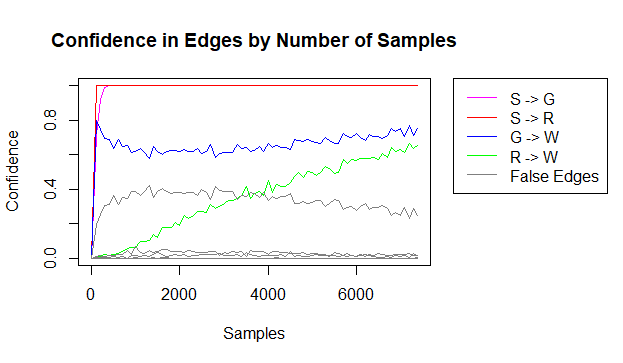}
  				\caption{Edge confidence for four-node network as a function of number of samples used to generate the BN.}
           		\label{fig:Confidence4Node}
			\end{figure}

			This variability is echoed in the edge confidence graph as shown in Figure~\ref{fig:Confidence4Node}, where we can see that an incorrect edge was predicted with higher frequency than one of the actual edges for low to medium sample counts.
			
			An interesting observation is that the collider $$G \rightarrow W \leftarrow R$$ is not picked up easily. If incorrectly constructed, any alterations to the collider would have relatively high effects on the SID, as intervention on the central node has a large impact on the interventional distribution. This could explains the variability seen in Figure \ref{fig:SIDSamplesFourNode}.
            
            \subsubsection{A Real Signalling and Regulatory Gene Network}

\begin{figure}[H]
           		\sidecaption
                 \includegraphics[scale=0.28]{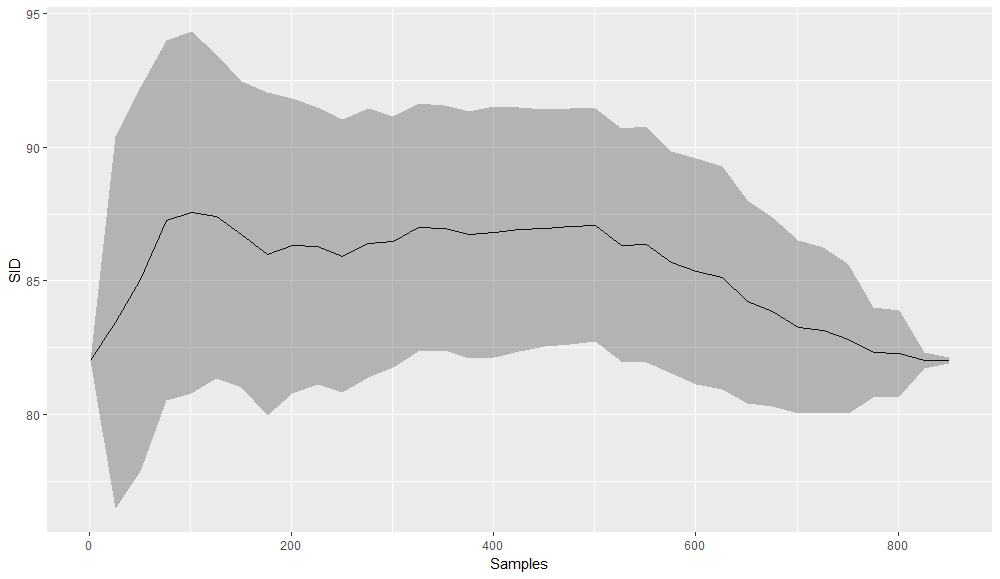}
  				\caption{Mean SID between the reference Sachs network and 300 BNs reconstructed from the samples, plotted as a function of the number of samples used to reconstruct the BN. $ 853 $ is the maximum available sample size. A grey band of +-1 standard deviation for the mean SID is shown.}
           		\label{fig:SIDsachs}
			\end{figure}

\begin{figure}[H]
\centering
\includegraphics[width = \textwidth]{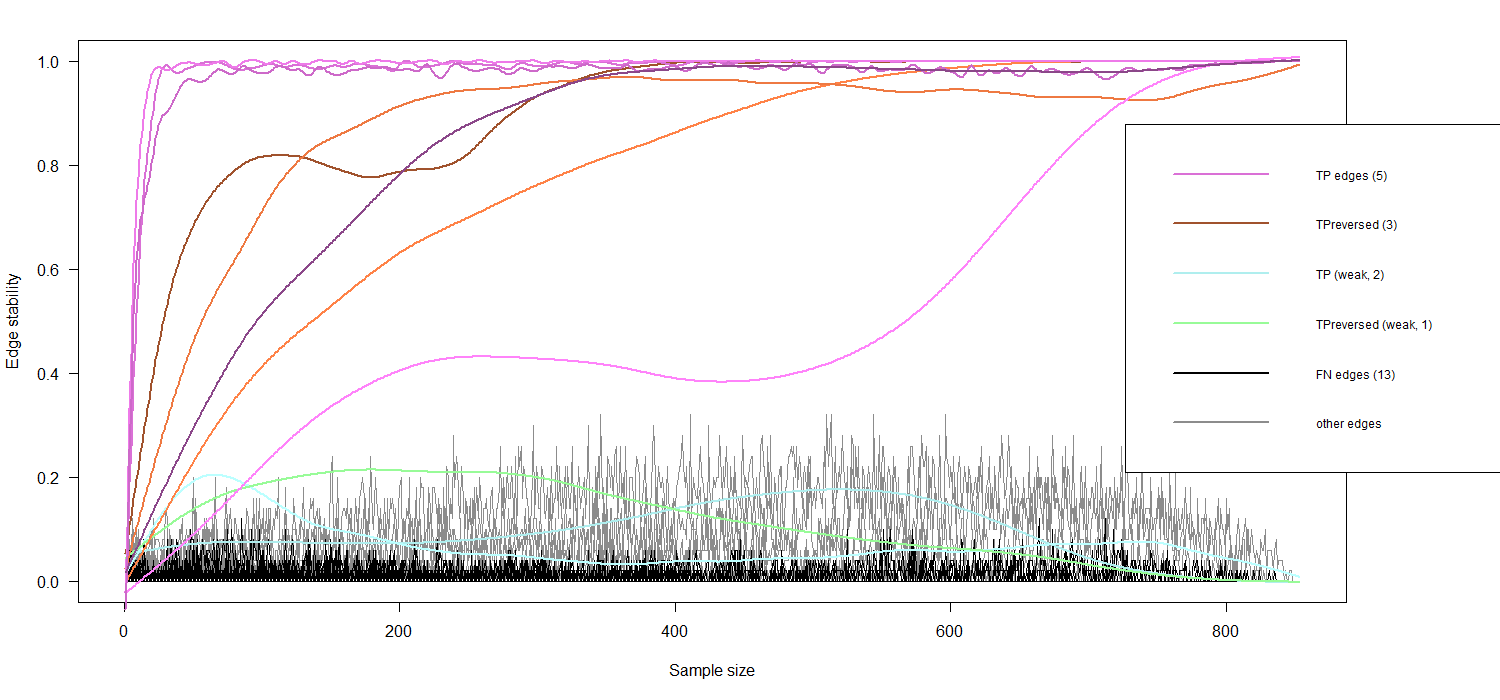}
\caption{\label{fig:conf-sachs}Sachs edge confidence as a function of sample size. Edge colours match those of Figure~\ref{fig:resu-sachs}.}
\end{figure}

\begin{figure}[H]
\sidecaption
\includegraphics[width=0.52\textwidth]{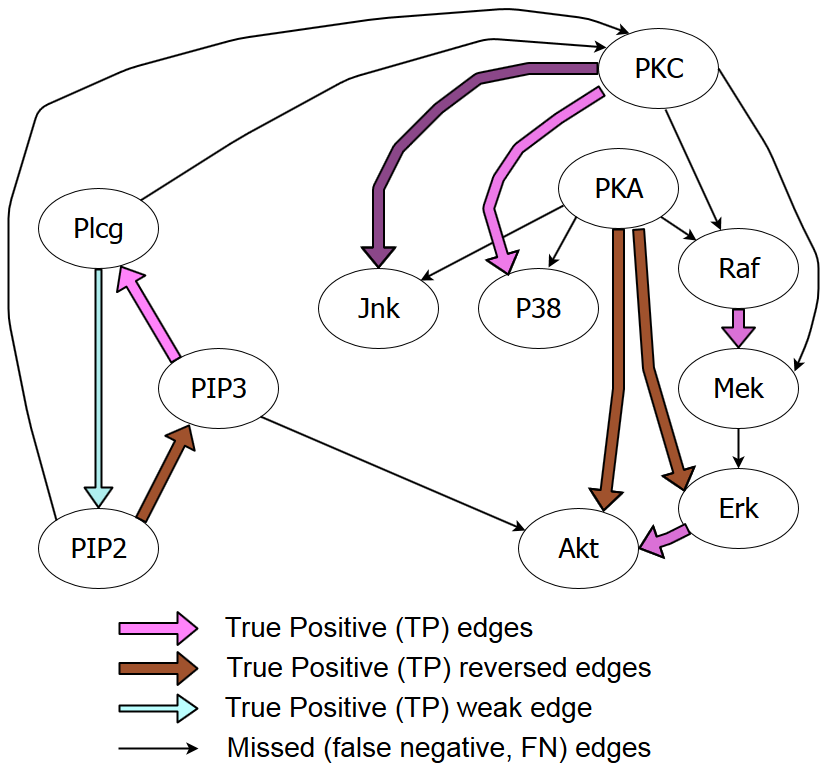}
\caption{\label{fig:resu-sachs}Final predicted network with our causal Bayesian network approach. Violet edges are correctly identified, pale blue are those with weak confidence, brown edges are those which were learned in the reverse direction compared to our reference network, and the thin black edges are those which were missed. Notice that one weak true positive edge and one weak, reversed, true positive edge are not represented (see Figure~\ref{fig:conf-sachs}) as they overlap edges represented in the opposite direction.}
\end{figure}

As an illustration, we applied the presented methodology to a real biological network. We chose the Sachs et al. 2005~\cite{sachs2005} 11-node network. We keep the same nomenclature for genes as the one used in the initial paper, except for Plc$ \gamma $, which is denoted Plcg here. We only used the $ 853 $ samples without intervention. The network which we considered as the reference network for SID comparison is depicted in Figure~\ref{fig:resu-sachs}. This network is ideal for our purpose as it has a large number of available samples, and does not include cycles. In addition, it has intervention data sets which could be used to refine the model. Results are presented in Figures~\ref{fig:SIDsachs}, \ref{fig:conf-sachs} and \ref{fig:resu-sachs}. 

The SID performance evolution in Figure~\ref{fig:SIDsachs} can seem a bit disappointing. This seemingly poor performance stems from the fact that as many edges are missing from the reconstructed network, the resulting graph finally reaches three disconnected components (PIP2-PIP3-Plcg, Raf-Mek and PKC-PKA-Jnk-P38-Akt-Erk). These disconnected components leave many pairs for which the interventional distribution is not adequately estimated, and hence increase SID~\cite{SID}. We see that adding the first edges even increases the computed SID. Forming a more complete sketch of the reference network helps to decrease the SID with increasing sample size. Notice that the empty network (predicted with no data) and the final 8-node network have the same SID measure, $ 84 $, although the predicted network is much more satisfying as it comprises 8 edges: five true positives and 3 reversed edges.

Figure~\ref{fig:conf-sachs} presents the edge confidence as a function of sample size. All true positive edges are identified very quickly (with less than 50 samples), except one (PIP3$ \to $Plcg) which requires more samples (more than 100) to stand out, becoming stable with more than $ 500 $ samples. This may be because it competes with another reversed edge (the green lines in Figure~\ref{fig:conf-sachs}). Furthermore, it is not clear whether this edge is indeed in this direction or reversed (see discussion in~\cite{sachs2005}). Another predicted edge assigned the wrong direction is also competing with the correct edge as per the reference network (PIP3$ \to $PIP2), and this was reported as bidirectional in the literature study of~\cite{sachs2005}. Note that the local PIP2$ \to $PIP3$ \to $Plcg$ \to $PIP2 loop does not contradict the DAG assumption, as this representation is that of the average graph created at that number of samples. The true positive edge with weak confidence is not part of the finally inferred network. When it is in the model, PIP2$ \to $PIP3 is either not present, or is reversed (hence in the correct direction, as per discussion above). 
The confidence of missed edges do not stand out from the background noise, i.e. the non-relevant edges or true negatives. Either the local encoded dependency is already captured by the $ 8 $ edges, or it requires interventional data to precisely identify them.

Five edges were correctly retrieved, and three additional edges were retrieved in the opposite direction compared to our reference network. This leads to a directed precision of $ 62.5 \% $ (undirected precision $ 100 \% $) and a recall of $ 28 \% $ (undirected recall $ 44 \% $). To be noted is that no false positive edge is predicted.
No interventional data was used, while \cite{sachs2005} used all the 5400 available samples with several subsets (6) of interventions.

\section{Conclusion / discussion} 
\label{conclu}

In this chapter, we discussed networks as a powerful approach to modelling complex living systems. We focused on gene regulatory networks, and their reconstruction from observational (aka wild-type) gene transcript data. We reviewed some methods dealing with the concept of causality, and we specifically focused on Bayesian Networks. We introduced BNs in detail, and tested their capacity to retrieve causal links on two toy network examples. Unexpectedly, the two experiments lead to different conclusions. While the learning inference went quite well from a causal relationship perspective for the seven-node network, it showed mixed performance in the four-node network. We were not able to see any trend in classification of directed edges which are correctly predicted versus those which are not correctly predicted. In the four-node network, edges originating from the source node $ S $ were detected with small sample size without any ambiguity, while it would not have been surprising if the learning algorithm had inverted one and not the other (to avoid creating a new V-structure). In addition, the converging V-structure to the sink node $ W $ was much more difficult to retrieve despite relatively large sample sizes (up to 5,000 samples) and contrasting simulated distributions. The conclusion is quite different in our seven-node network, with all edges but one retrieved without ambiguity with more than 500 samples. Again, the most difficult directed edge to retrieve was one involved in a V-structure, this time from a source node ($ B $). It also required fewer samples than the four-node network, while much more structures are possible with 7 nodes. At the same time, this seven-node network contains more conditional independence than the four-node network. Another surprising result is that edges which could have been inferred in one direction or the other ($ B \to D $ could be inverted, or the directed path $ B \to E \to G $ could be inverted; not both at the same time) were assigned a direction without ambiguity. The most surprising conclusion so far in our experiments is that increasing the sample size seemed to always improve the identification of edge, including its direction, whether it is imposed in the equivalence graph or not. This could be due to not studying enough graphical configurations. This conclusion must also be qualified in that obtaining a few hundred samples for a small 10-node example is not very realistic from an experimental point of view.

\paragraph{Future work}

Obviously, we only showed very limited experiments in very special cases. For our conclusions to be valid, we would need to extend our experimental setup to (i) a bigger variety of conditional distributions, (ii) more local network motifs e.g. feed-forward loops, (iii) question the binary variable framework, and (iv) test other learning inference methods. It would be interesting to test point (i) by checking some well chosen edges for their prediction confidence as a function of the conditional probabilities, with a similar representation to that in Figure~\ref{fig:EdgeConfidence}, but a variation on the conditional distribution(s) on the x-axis instead of the sample size. However, varying a conditional distribution with one parameter in one dimension is technically challenging and necessarily not perfect. For point (ii), a next step would be to build a dictionary gathering the typology of local structures which can be encountered in networks and how well a BN learning algorithm could perform at retrieved directed edges on those structures. For point (iv), we and other colleagues~\cite{geurts2013} noted for example that random forests seem to give relevant directions to edges when used to infer GRN (see~\cite{huynh-thu2010}).

When performing tests to be presented in this chapter, we also used the graph edit and the KL divergence as potential measure of closeness between the true network and the predicted network. It is known that the graph edit distance is not ideal as two graphs with only one edge difference can lead to quite different causal conclusions and/or independence relationships. The KL divergence proved to be useful to compare probability distributions, yet prohibitively difficult to generalise. Combining KL-Divergence with the SID metric could provide an alternative representation of the ability of the learning algorithm to retrieve the distribution - if not the causal relationships. An easily generalised metric combining both, or some variation of the two could be of particular interest.


\section*{Acknowledgement}

Alex White was partly supported by a Summer Scholarship grant from the Institute of Fundamental Sciences at Massey University (NZ). This work was also supported by a visiting professor scholarship from Aix-Marseille University granted to Matthieu Vignes in 2017. 
The material in this chapter slowly hatched after many discussions with a multitude of quantitative field colleagues and biologists during some works we were involved in, e.g. \cite{vignes2011, vandel2012, marchand2014}, but most of our inspiration stems from other works we read about and discussed (e.g. \cite{chiquet2009, huynh-thu2010, rau2013, gagneur2013, vallat2013} to cite only a few). We are very appreciative of the many discussions with Stephen Marsland (Victoria University of Wellington, NZ).
Lastly, we are very grateful to reviewers of this chapter for their insightful comments.

\bibliography{bibliography}
\bibliographystyle{plain}

\end{document}